# A Punctuated Equilibrium Analysis of the Climate Evolution of Cenozoic exhibits a Hierarchy of Abrupt Transitions

Denis-Didier Rousseau[1,2,3 *], Witold Bagniewski[4], Valerio Lucarini[5,6,7]

[1] Université Montpellier, Géosciences Montpellier, Montpellier, France
[2] Silesian University of Technology, Institute of Physics-CSE, Division of Geochronology and Environmental Isotopes, Gliwice, Poland
[3] Columbia University, Lamont Doherty Earth Observatory, New York, USA
[4] Ecole Normale Supérieure – Paris Sciences et Lettres, Laboratoire de Météorologie Dynamique, Paris, France
[5] University of Reading, Department of Mathematics and Statistics, Reading,
[6] University of Reading, Centre for the Mathematics of Planet Earth, Reading, UK
[7] School of Systems Science, Beijing Normal University, Beijing, PRC

(*) corresponding author: denis-didier.rousseau@umontpellier.fr.

**Abstract**

The Earth's climate has experienced numerous critical transitions during its history, which have often been accompanied by massive and rapid changes in the biosphere. Such transitions are evidenced in various proxy records covering different timescales. The goal is then to identify, date, characterize, and rank past critical transitions in terms of importance, thus possibly yielding a more thorough perspective on climatic history. To illustrate such an approach, which is inspired by the punctuated equilibrium perspective on the theory of evolution, we have analyzed 2 key high-resolution datasets: the CENOGRID marine compilation (past 66 Myr), and North Atlantic U1308 record (past 3.3 Myr). By combining recurrence analysis of the individual time series with a multivariate representation of the system based on the theory of the quasi-potential, we identify the key abrupt transitions associated with major regime changes that separate various clusters of climate variability. This allows interpreting the time-evolution of the system as a trajectory taking place in a dynamical landscape, whose multiscale features describe a hierarchy of metastable states and associated tipping points.

Early evidence of abrupt transitions in Camp Century and Dye 3 Greenland ice cores[1,2] attracted a lot of attention from the paleoclimatic community before being well acknowledged and understood. These findings gave evidence of a sequence of previously unknown abrupt climatic variations. However, such transitions did not seem to agree with other marine and terrestrial records, which led to considerable debate in the field[3–5]. Over the course of several decades spent retrieving and studying more detailed paleorecords, the existence of these rapid climatic variations, known as Dansgaard-Oeschger events (DO), has become well accepted. Further support for these findings has recently been provided by the identification of additional abrupt transitions from the NGRIP ice core, which has been made possible by the increased temporal resolution of the record[6]. These additional events correspond to changes of either short duration or amplitude in the stable $O^{18}$ over $O^{16}$ isotopes ratio $\delta^{18}O$ that visual or standard statistical inspections do not necessarily flag. We remind that the fractionation of oxygen isotopes can be used to reconstruct local temperature conditions.

The Earth climate has experienced numerous abrupt and critical transitions during its long history, well beyond the specific examples above[7,8]. As discussed in Rothman [9] following Newell [10], rapid variations change are more likely to lead to catastrophic consequences in the biosphere - as in the case extreme case of mass extinction events - because it is hard for the evolutionary process to keep pace with shifting environmental conditions. Such transitions are often referred to as climatic tipping points (TPs), associated with possibly irreversible changes in the state of the system. The term TP was originally introduced in social sciences[11] and made popular more recently by Gladwell[12]. The study of TPs has recently gained broad interest and perspective in Earth and Environmental sciences, especially with regard to the future of our societies under the present climate warming scenarios[13–16]. The term Tipping Elements (TE) was introduced in[15,17], and subsequently adopted by other authors[14,18] to characterize the specific components of the Earth System that are likely to experience a TP in the near future as a result of the ongoing climate crisis[16,19–21]. Recently, the concept of TP has been used to define, in turn, rapid societal changes that might lead to potentially positive impacts in terms of addressing the climate crisis[22,23].

Here, we want to investigate critical transitions in the Earth climate history by looking at two key high-resolution datasets that show evidence of abrupt transitions. The first dataset is the CENOGRID benthic $\delta^{18}O$ and $\delta^{13}C$ record corresponding to the compilation of 14 marine records over the past 66 Myr, from the Cretaceous-Paleogene (K–Pg) extinction event till present[24]. The second dataset comprises the North Atlantic U1308 benthic $\delta^{18}O$, $\delta^{13}C$ and $\delta^{18}O$ bulk carbonate time series covering the past 3.3 Myr[25].

While visual evidences of abrupt transitions have already been discussed for these datasets, we wish to identify key abrupt thresholds by applying the recurrence quantification analysis (RQA) to each individual univariate time series and supplementing it with the Kolmogorov-Smirnov (KS) test[26], see[27] and discussion below. Then, the selected transitions are discussed in the context of the Earth climate history allowing the definition of dynamical succession of abrupt transitions. Such transitions are then interpreted taking into account the evolution of key climate factors such as $CO_2$ concentration, average global sea level, and depth of the carbonate compensation.

The existence of TPs is intimately related to the multistability properties of the climate system, which have long been recognised in different contexts; see e.g.[28–31] and discussion in[32,33]. The multistability of the climate system comes the presence of more than one possible climate states for a given set of boundary conditions[34]. While earlier analyses have mostly evidenced the possibility of bistable behaviour, multistability can indeed include multiple competing states[35–38]. Recently, it has been proposed that the metastability properties of the climate system can be understood by interpreting the climate evolution as a diffusion process taking place in an effective dynamical landscape[37,39,40] defined by the the Graham's quasi-potential[41]. The local minima of the quasi-potential indicate the competing metastable states, with the transitions between such states occurring preferentially through the saddles of the quasi-potential. Such a viewpoint mirrors earlier proposals for interpreting biological evolution, namely the Waddington's epigenetic landscape[42–46], see also relevant literature associated with synthetic evolution models like Tangled Nature[49,50], and foresee the climatic transitions associated with the TPs as a manifestation of a dynamics characterized by punctuated equilibria[47-50,] The interplay between periods of stasis and rare transitions between competing metastable states seem to reflect the fact that climate fluctuations behave according to the dominance of stabilizing vs destabilizing feedbacks when considering short vs long time scales, respectively[51]. We will see that in most cases the RQA applied to the CENOGRID $\delta^{18}O$ dataset flags TPs that disagree in terms of dating from those derived from the $\delta^{13}C$ dataset. This is unsurprising because the two proxy data are sensitive to vastly different climatic processes. Nonetheless, the candidate TPs from both datasets come hand in hand with saddles of the bidimensional quasi-potential estimated from the bivariate time series. This indicates consistency between separate ways of detecting TPs. Additionally, TPs featuring faster characteristic time scales are associated with smaller-scale decoration of the quasi-potential, in agreement with what conjectured in[37]. The analysis of the benthic $\delta^{18}O$ and $\delta^{13}C$ suggests that the evolution of the climate in Cenozoic is characterized by a hierarchy of TPs due to an underlying multiscale quasi-potential.

## Results

### 1. Detecting Critical Transitions of the past 66 Myr – 3 Myr history of the Earth Climate

The augmented KS test of the benthic $\delta^{18}O$ record of the past 66 Ma identifies six major abrupt transitions corresponding to two major warming events at about 58 Ma and 56 Ma, followed by four major coolings at 47 Ma, 34 Ma, 14 Ma and 2.8 Ma respectively (Fig. 1A). The competing metastable states associated with these transitions feature rather long temporal persistence (a long time-window of 1-4 Ma is used, see Suppl. Mat.). These events are classical ones described from the literature[52], where the first two transitions led to warmer conditions, while the latter four led to colder conditions. The same transitions are identified by employing the recurrence plot (RP) and recurrence rate (RR) analyses[27,53], which also identify three more events, occurring at around 63 Ma, 40 Ma and 9.7 Ma, see Fig. 1C, Suppl. Tab. 1. We have chronologically labeled these $TP_O1$ to $TP_O9$, where the lower index refers to the used proxy data. As shown in Fig 1B, TP$_O$6 separates the climate variability in two separate macroclusters prior and after 34 Ma, the well-known Eocene-Oligocene Transition (EOT) [54], which is a key step in the Cenozoic climate history and is associated with a major extinction event[55,56].

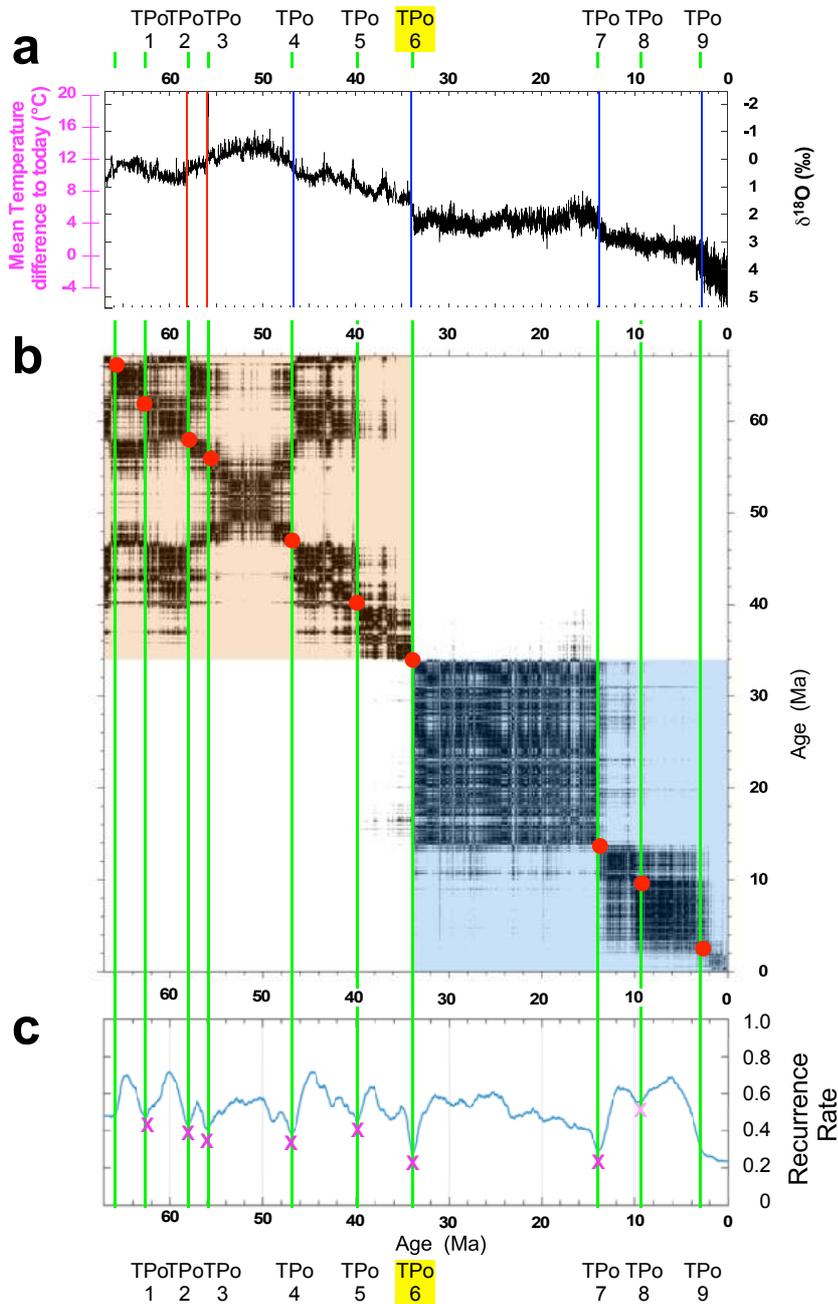

**Figure. 1.** KS test and Recurrence Quantification Analysis (RQA) of CENOGRID benthic δ$^{18}$O. a) Time series in Ma BP with difference of the reconstructed and present Mean Global Temperature in pink). KS test identifying abrupt transitions towards warmer conditions in red and cooler or colder conditions in blue; b) Recurrence plot (RP) with identification of the main two clusters prior and after 34 Ma. The main abrupt transitions identified are highlighted by red circles, and c) Recurrence rate (RR). The pink crosses and vertical green lines indicate the abrupt transitions (TP$_O$) detected by the RQA. CENOGRID benthic δ$^{18}$O data are from Westerhold et al. [24]

The older macrocluster corresponds to very warm climate conditions and, using Marwan et al.'s nomenclature[53], a disrupted variability. The average global temperature was estimated to be between 8°C and 16°C above the present day one, with no apparent presence of any major continental ice bodies[24] (Fig. 1A). Five major transitions are found within this period. They include the 2 major abrupt warmings at 58 Ma (TP$_O$2) and 56 Ma (TP$_O$3). These two transitions are thresholds towards much warmer oceanic deep water characterizing the first late Paleocene-Eocene hyperthermal[57,58] and Paleocene-Eocene Thermal Maximum (PETM)[59,60] respectively. The third one at about 47 Ma (TP$_O$3) represents a transition towards cooler deep waters and named the Early-Middle Eocene cooling[58]. The last one at about 40 Ma (TP$_O$4) inaugurates the period of continuous cooling due to the decrease of $CO_2$ concentrations that eventually leads to the TP$_O$6 event[61], which is associated with the build-up of the Antarctica ice sheets. TP$_O$6 separates two fundamentally different modes of operation of the climate system, corresponding, as discussed later in Sect. 4, to two separate minima of the quasi—potential. From 34 Ma to present day the records feature more positive values of benthic $\delta^{18}O$ associated with prevailing colder climate conditions. After TP$_O$6, the climate featured mostly stationary conditions with a slight warming until the Middle Miocene Climate Transition (TP$_O$7) that occurred around 14 Ma[62,63]. The last major transition (TP$_O$9) occurred around 2.8 Ma, leading to the Pleistocene and the onset of glaciations in the Northern Hemisphere.

What discussed above provides a more systematic and robust counterpart of the analysis of TPs presented in[24]. In general, we prioritize the information coming from the $\delta^{18}O$ record because of its very strong link to the Earth's temperature. Nonetheless, the analysis of the Benthic $\delta^{13}C$ record performed along the same lines as above provides separate pieces of information on the critical transitions of the Earth's Climate. Benthic $\delta^{13}C$ values characterize deep-water ventilation with high $\delta^{13}C$ values in regions close to deep-water formation area. The KS analysis performed over a time window of 1-4 Myr individuates 14 TPs, with the RP suggesting an additional one, located at around 34 Ma and associated with the EOT. We refer to these 15 TPs associated with the $\delta^{13}C$ record as TP$_C$ 1 to 15; see Suppl. Fig. 1, Suppl. Tab. 1. The interval 56.15 Ma – 7.15 Ma is well characterized, showing some subclusters distributed around 34 Ma. The 56.15 Ma date groups $\delta^{13}C$ values above 1‰, at the base of the record, while 7.15 Ma gathers the negative $\delta^{13}C$ values which mainly occurs at the top of the record. The two periods are characterized by very different climatic conditions. The earlier climate regime features more input of carbon in the ocean while the later climate, instead, is characterized by higher presence of carbon in the atmosphere.

## 1.1 Discussion of the Detected Critical Transitions

We next want to analyze TP$_O$ 1-9 in relation to different reconstructed paleoclimatic data, namely the global mean sea level (GMSL), the Pacific carbonate compensation depth (CCD), and $CO_2$ concentration (Fig. 2). Using benthic foram $\delta^{18}O$ and Mg/Ca records from high-resolution Pacific cores which were not included in the CENOGRID compilation, Miller et al.[58] have reconstructed variations in the GMSL over the past 66 Ma. By measuring the carbonate content in Pacific sediment cores and applying transfer functions, Pälike et al.[64] have generated a detailed Cenozoic record of the Pacific CCD, which denotes the depth below which carbonates dissolve. Finally, by compiling estimates from various proxies including foram $\delta^{13}C$, boron isotopes, stomata, paleosols, Beerling & Royer[65] have produced a comprehensive Cenozoic record of the $CO_2$ concentration. The signature of the TP$_O$6 is evident in the three records, corresponding to an abrupt decrease of the GMSL by about 70m and of the CCD by around 1000m.

Additionally, TP$_O$6 marks the start of a progressive decrease in CO$_2$ levels, from approximately 750 ppm to values of the order of 280 ppm.

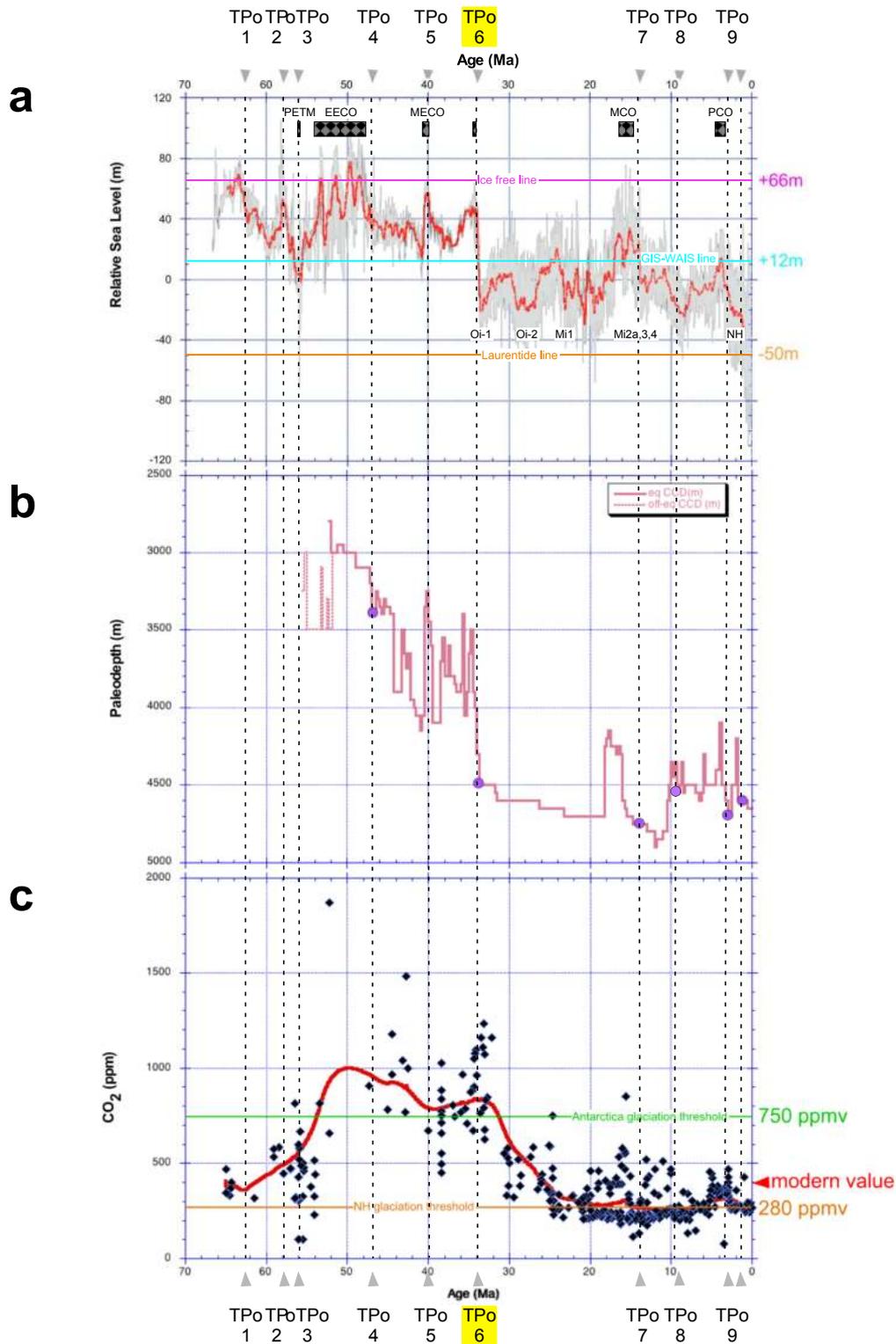

**Figure 2.** Variation through time of three main climate factors and comparison with the identified abrupt transitions (TP) in the CENOGRID benthic δ$^{18}$O. a) Global Mean Sea Level in meters from Miller et al. [58]. Identification of specific warm and of glaciation events. The Laurentide, GIW-WAIS and Ice free lines are from Miller et al. [58]; b) Carbonate Compensation Depth (CCD) in meters from Pälike et al. [64]. The purple circles identify the TPs on this record; c) Estimate of the CO$_2$ concentration in parts per million volume (ppmv) from Beerling & Royer[65]. The Antarctica glaciation threshold at 750 ppmv and the NH glaciation threshold at 280 ppmv lines are from DeConto et al. [66]

Along the lines of Westerhold et al. [24], one can identify four competing states,"Warmhouse" (66 Ma-TP$_O$1 and TP$_O$3- TP$_O$5) and "Hothouse" (TP$_O$1-TP$_O$3) climates in the earlier, warmer period, followed by the "Coolhouse" ($TP_O5$-$TP_O6$) and "Icehouse" climates ($TP_O6$ to present); see Fig. 1. The first two states alternated in a warm-hot-warm sequence under extremely high $CO_2$ concentrations[65] as compared to those measured over the past 800 Kyr in the Antarctic ice cores, which represent the reference states for the IPCC potential scenarios of climate change warming[67] (Fig. 2C). Before 34 Ma, one finds substantially larger values for GMSL, CCD depth, and $CO_2$ concentration whose average values are: +38 ± 15 m, 4600 ± 150 m, and 630 ± 300 ppmv respectively (Suppl. Tab. 2). Note that the $CO_2$ concentration is much higher than the levels observed over the past 800 kyr in the Antarctic ice cores, which represent the reference states for the potential scenarios of climate warming outlined by the IPCC[67]. Conversely, from 34 Ma until present, the Earth experienced much lower $CO_2$ concentrations and GMSL, thus generating the classical climate trend towards the recent ice-age conditions[8,24,52] (Fig. 2A,C). Indeed, the last 34 Myr show average values in GMSL, $CO_2$ concentration, and CCD depth of -3.5 ± 13 m, 330 ± 160 ppmv, and 3500 ± 400 m, respectively (Suppl. Tab. 2), which are much lower than in the older interval. This second set of means has underestimated values since it is based on data from the Miller et al.[58] dataset, which ends at 0.9 Ma,

These GMSL, CCD and $CO_2$ reconstructions show key transitions that fit with the CENOGRID thresholds deduced from the KS and RR analysis of the benthic $\delta^{18}O$ as they signal an increase or a decrease in the global mean sea level corresponding roughly to warming or cooling episodes of the Earth history or strong variations in the concentration of atmospheric $CO_2$. The variations observed in $CO_2$, CCD and GMSL at the 9 identified TPs from the benthic $\delta^{18}O$ record indicate heterogeneous characteristics prior to the TP6 major threshold (Suppl. Tab. 3). On the contrary, more homogenous features are noticed in the three climate proxies after $TP_O6$, translating the occurrence of major reorganizations in the climate system, which become interesting to test at shorter timescales; see discussion below.

**2. Quasi-potential Landscape and Critical Transitions**

As discussed in detail in the Methods section, taking inspiration from the use of the Waddington epigenetic landscape to describe evolution[42–46] and from the theory of punctuated equilibrium[49,50], it has been proposed in[37,39,40] to study the global stability properties of the climate system by introducing an effective quasi-potential[41], which generalizes the classical energy landscape framework often used to study metastable stochastic system. The quasi-potential formalism allows one to study general non-equilibrium system and to associate local maxima of the probability distribution function (pdf) of the system stable states. Additionally, saddles of the pdf are associated with Melancholia (M) states[34], which are unstable states living in the boundary between different basins of attraction. In the weak-noise limit, noise-induced transitions between competing stable states are expected to go through such M states. What we have done here is to construct the empirical bivariate pdf of climate system in the projected ($\delta^{13}C$, $\delta^{18}O$) and check whether $TP_O$ 1-9 indicated in Fig. 1 correspond, at least approximately, to saddles of the pdf. We find – see Fig 3A – that, indeed, this is the case for $TP_{OS}$ 1, 2, 4, 5, 6, 8, 9, whereas no agreement is found for $TP_{OS}$ 3 and 7, which seem to take place in regions where the density is very small and very large, respectively. A very similar version of the bivariate pdf shown here had already been used in Westerhold et al.[24] to identify the Icehouse, the Coolhouse, the Warmhouse, and the Hothouse states, as groupings of nearby peaks corresponding to qualitatively similar climates.
A major improvement we propose here is to identify the transitions between such states, as well as less obvious transitions, by looking at the saddle points. TPo6 clearly emerges as the most important transitions, as it basically breaks the pdf into two separate parts. Note that the transitions shown in Fig. 1 are associated, because of the choice of a long time-window, with events that occur over long periods and that lead to persistent changes in the state of the system.

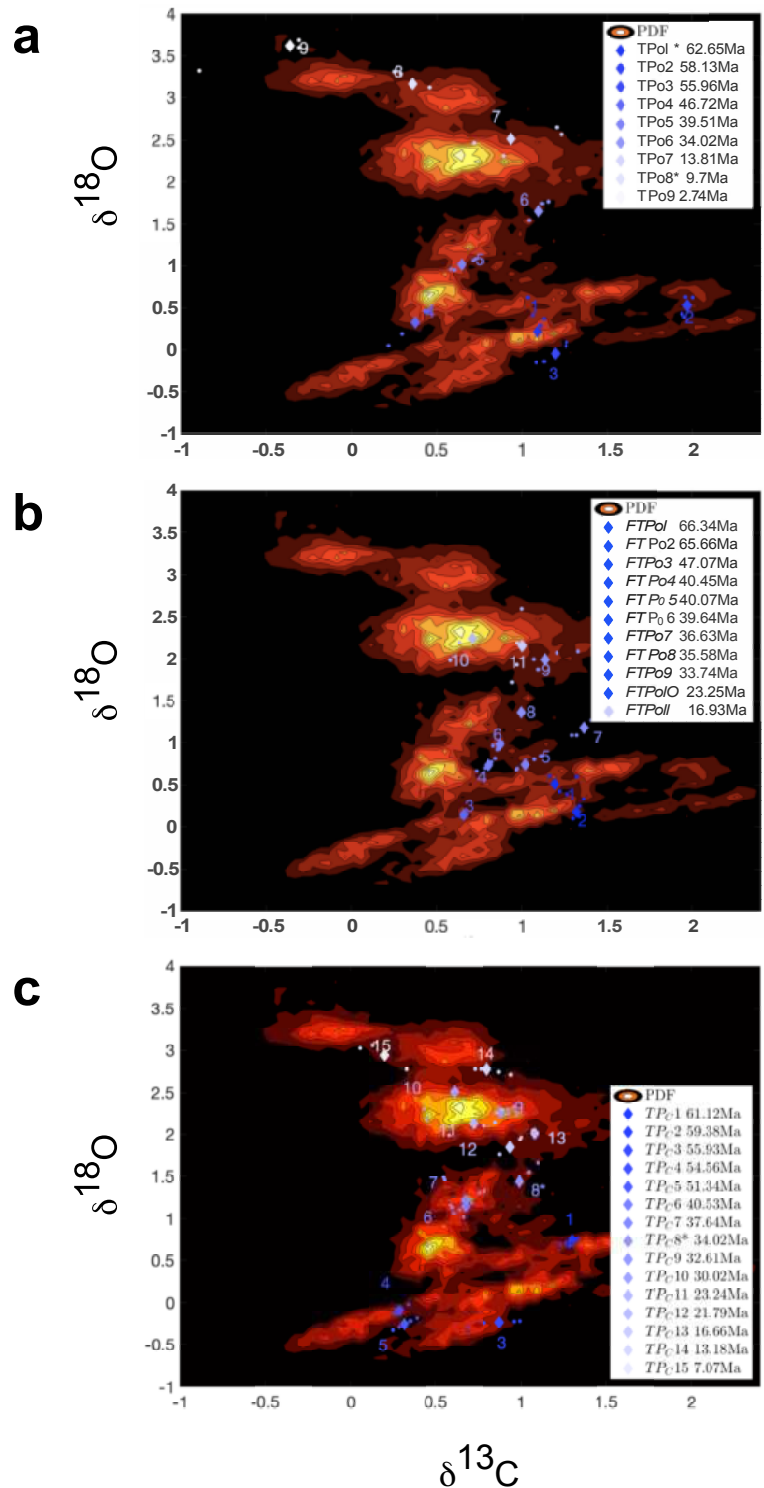

**Figure 3**. Probability density of the climate system in the projected CENOGRID benthic $\delta^{18}O$ and $\delta^{13}C$ space. a) Chronologically ordered TP$_O$s (diamonds) selected according to the KS methodology for $\delta^{18}O$ with time window 1-4 My are shown. The two extra TP$_O$s found via RP are indicated with a *. b) Chronologically ordered fast TP$_O$s (FTP$_O$s, diamonds) selected according to the KS methodology for $\delta^{18}O$

with time window 0.25-1 Myr are shown. c) Same as a), but for the $\delta^{13}$C record. The approximate timing of the TPs is indicated (rounded to .01 My). The 5 Ky-long portions of trajectories before and after each TP are also plotted.

By combining the recurrence analysis with the quasi-potential formalism we can extract further relevant information. It is natural to ask ourselves what happens if, instead, we consider a catalogue of transitions for the $\delta^{18}$O record that are detected by considering shorter time windows (0.25 Ma-1Ma) in the KS procedure. One finds – see Fig. 3B - 11 of such transitions (see Suppl. Mat.). Once we report such fast TP$_O$s (FTP$_O$s) into the empirical bivariate pdf of the climate system in the projected ($\delta^{13}$C, $\delta^{18}$O) space, we find that they correspond to finer and smaller structures of the pdf as opposed to the case of the TPs. Hence, events that are associated with faster time scales are associated with smaller jumps between secondary maxima in the pdf belong to a hierarchically lower rung than those occurring over longer time scales. This seems to support the proposal made in[37].

As additional step, we repeat the same analysis leading to Fig 3A by using, instead, the $\delta^{13}$C record, which features, as mentioned above, a total of 15 TP$_C$s. Figure 3C shows clearly that the TP$_C$s are, for the most part, saddles that had not been flagged by the TP$_O$s. We have evidence that the same saddle is crossed more than once in a back-and-forth fashion few millions of years apart (e.g. TP$_C$ s 4 & 5; TP$_C$s 6 & 7; TP$_C$s 12 & 13), which supports the dynamical interpretation discussed here. Comparing Figs 3A and 3C, one discovers that the same saddle is responsible for TP$_O$8 and for TP$_C$15 even if the dating is different. Similarly, the same saddle is responsible for TP$_O$5 and TP$_C$s 6 & 7. Two key climatic features that appear as TPs in both records: the PETM is analogous to TP3 in both records, while TP$_O$6 and TP$_C$8 represent the EOT.

**3. The recent past**

The past 3.3 Myr record from North Atlantic core U1308 can be considered as a blow-up of the CENOGRID dataset (Fig. 4). As previously mentioned, the last 3.3 Myr have been defined as an Icehouse climate state, with the appearance, development, and variations of the NHIS[24], whilst the Antarctic ice sheets had already mostly reached their maximal expansion. The variations in the deep-water temperature, as expressed by the benthic $\delta^{18}$O, are interpreted as an indicator of the continental ice volume with clear interglacial-glacial successions[68–70]. The Icehouse state is characterized by a change of the interplay between benthic $\delta^{13}$C and $\delta^{18}$O, which corresponds to a new relationship between the carbon cycle and climate[71]. Indeed, one finds a very strong correlation between the two records (Pearson's coefficient being approximately -0.6). The correlation mainly results from the fact that the time series have approximately a common quasi-periodic behavior due to amplified response to the astronomical forcing as dictated by the Milankovich theory. Note that that the presence of such almost regular resonant oscillations makes the use of the quasi-potential framework not particularly useful for describing the dynamics of the system, so that we will not pursue this approach for the analysis of the Quaternary records.

The KS augmented test and the RQ of the benthic $\delta^{18}$O agree in identifying six abrupt transitions dated at approximately 2.93 Ma, at 2.52 Ma, at 1.51 Ma, at 1.25 Ma, at 0.61 Ma and at 0.35 Ma (Fig. 4, Suppl. Tab. 1). They characterize the dynamics of North Hemisphere ice sheets (elevation and spatial expansion) and agree with the classical transitions characterizing the Marine Isotope Stages (MIS) as already observed in numerous records covering the same interval (see Rousseau et al. [72] and references. therein).

The first two transitions broadly match the previously discussed CENOGRID's $TP_O9$ associated with the onset of the Pleistocene (note that the interval between the two is much smaller than the resolution needed to separate to $TP_O$s) and are followed by four more recent TPs associated with the benthic $\delta^{18}O$ record ($RTP_O$s). There is clear evidence of the Mid-Pleistocene (MPT) critical transition, between 1.25 Ma and 0.8 Ma, during which a shift occurred from climate cycles dominated by a 40-kyr periodicity (due to obliquity) to 100-Kyr periodicity (due to eccentricity) dominated ones[73–77]. The 1.25 Ma date is particularly significant, since it is followed by an Increase in the amplitude of glacial–interglacial fluctuations.

A complementary RQA of the $\delta^{18}O$ bulk carbonate record from U1308, which characterizes episodes of iceberg calving into the North Atlantic Ocean IRD released into the North Atlantic Ocean[25], and therefore illustrates the dynamics of the Northern Hemisphere ice sheets (NHIS), yields similar dates to those obtained for the benthic $\delta^{18}O$ record (see Suppl. Fig. 3, Suppl. Tab. 1) [72]. Indeed, one finds abrupt transitions at 2.75 Ma, at 1.5 Ma, at 1.25 Ma, at 0.9 Ma and 0.65 Ma. Finally, as opposed to the CENOGRID $\delta^{13}C$ RP, U1308 $\delta^{13}C$ RP shows a drifting pattern similar to that of benthic $\delta^{18}O$, with only 2 key transitions at 2.52 Ma and 0.48 Ma (see Suppl. Fig. 3). Note that the 0.48 Ma transition does not have any equivalent in the benthic $\delta^{18}O$ records.

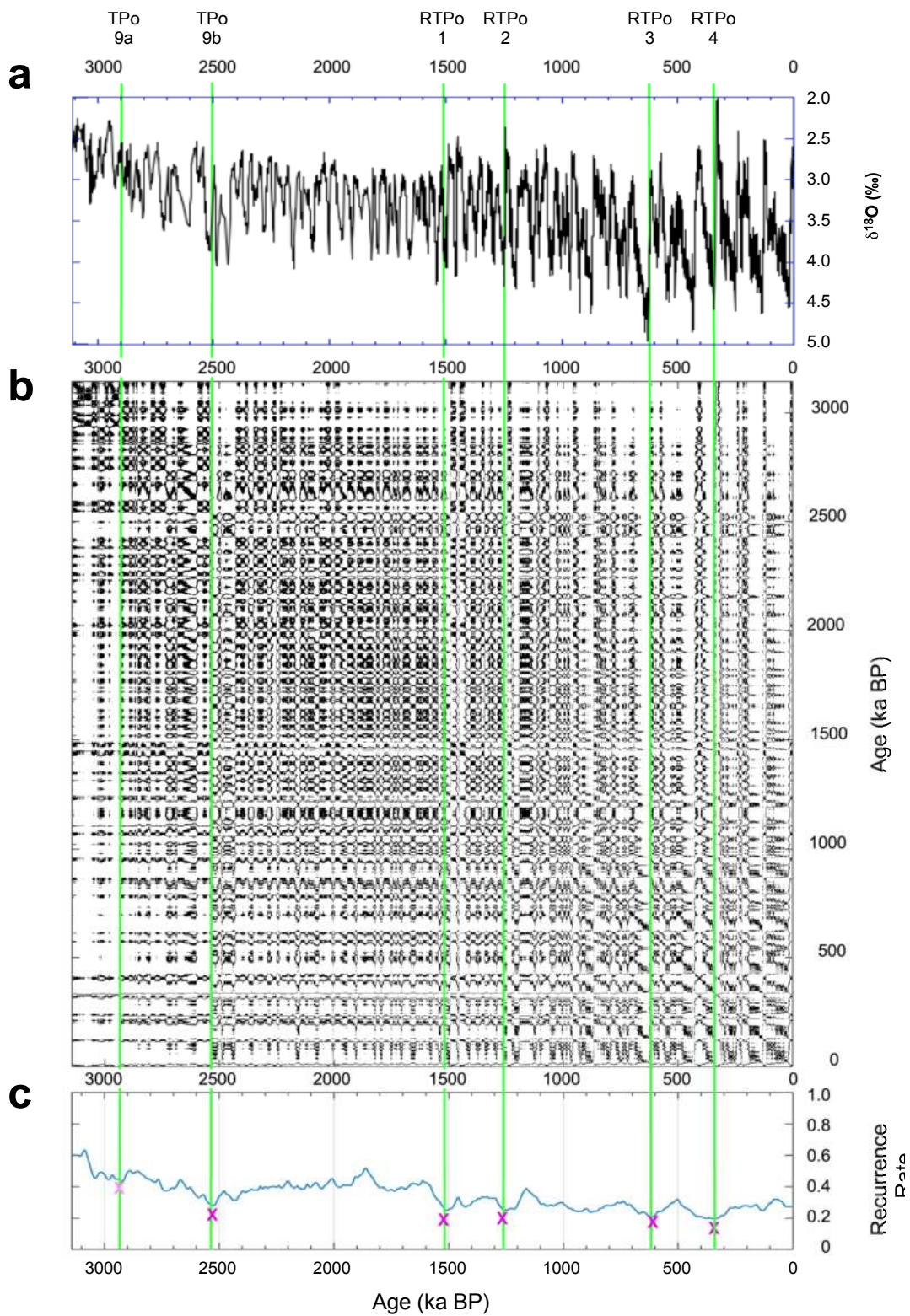

**Figure 4.** RQA of U1308 benthic $\delta^{18}O$. a) Time series in ka BP; b) RP; and c) RR. Pink crosses and green lines as in Fig. S1. $TP_O9$ and $RTP_O1$-4 abrupt transitions are identified from the RR. U1308 benthic $\delta^{18}O$ data are from Hodell & Channell[25]

## Discussion

Studying the same CENOGRID dataset, Boettner et al.[78] identified 9 geological transitions at respectively 62.1 Ma, 55.9 Ma, 33.9 Ma, 23.2 Ma, 13.8 Ma, 10.8 Ma and 7.6 Ma. Four of them are indeed identical to those determined in the present study: 62.1 Ma, 55.9 Ma, 33.9 Ma and 13.8 Ma, the first two being preceded by a significant early warning signal[78], which is instead absent in the case for the EOT key transition.

Based on the results of both the RQA and the KS test of the $\delta^{18}O$ and $\delta^{13}C$ time series considered in this study, and the bivariate analysis performed using the framework of the quasi-potential theory, we propose a succession of critical transitions as described in Fig. 5. The critical transitions $TP_O1$ to $TP_O9$ shaped the Earth climate towards the onset and development of the Southern ice sheets and the later build-up of the NHIS. $TP_O9$ is followed by four more recent RTPs during the Quaternary, which steered the evolution of the ice sheets and of the climate as a whole until the present day. The climatic evolution during the Cenozoic until about 3 Ma seems to conform to a punctuated equilibrium framework, where the $TP_OS$ are associated with rapid transitions between rather different metastable modes of operation of the climate system. In particular, a key step that separates two rather diverse sets of climatic states occurred around 34 Ma at the EOT ($TP_O6$). Without the major drop in GMSL, in $CO_2$ concentrations, and in CCD, the Earth climate could have been different. However, after $TP_O6$, the Earth climate entered new dynamical regimes marked by much lower $CO_2$ concentrations, a lower GMSL, and a lower CCD. The remodeling of oceanic basins and mountain uplifts changed the marine and atmospheric circulations patterns, which played a crucial role in initiating and shaping the development of the NHIS.

Interestingly, the analysis of the $\delta^{13}C$ time series identifies a different set of critical transitions, with the exception of the PET ($TP_O2$ and $TP_CC$) and the EOT ($TP_O6$ and $TP_C8$), which are identified for both analyzed proxies. Looking into the bivariate pdf in the projected ($\delta^{13}C$, $\delta^{18}O$) space allows a better understanding of the nature and the origin of the TPs separately detected by studying the recurrence properties of the univariate time series. Indeed, we are in most cases able to associate both the $TP_OS$ and the $TP_CS$ to transitions across saddles of the effective quasi-potential. Clearly, some critical transitions might be more easily detectable when examining one time series rather than the other, because different proxies might be more sensitive to the active climatic processes, yet the approach taken here allows placing all TPs within a common ground. Additionally, TPs associated with faster processes and occurring between slower TPs correspond to transitions across smaller-scale saddles of the quasi-potential, hence revealing an analogy between the hierarchy of TPs and the multiscale nature of the quasi-potential.

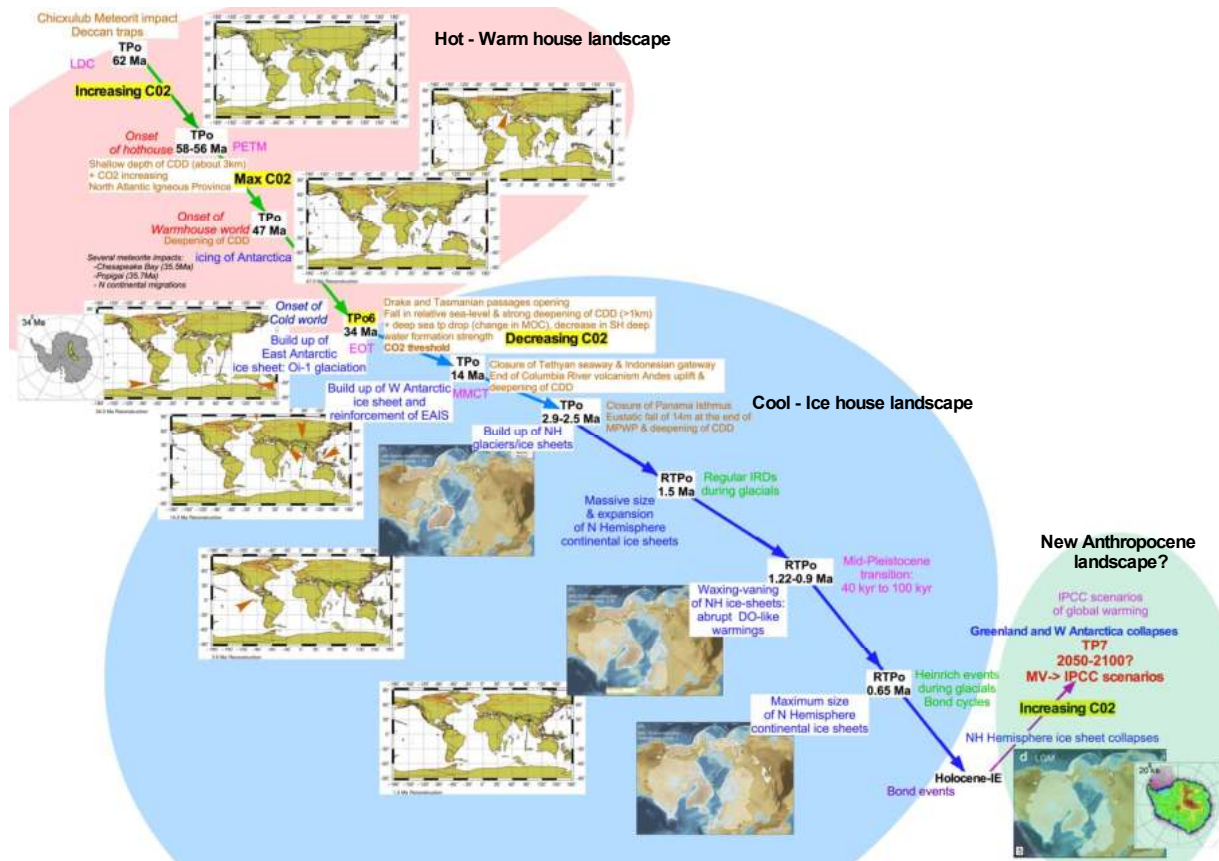

**Figure 5.** Evolution of the Earth Climate history among 2 different dynamical landscapes and a potential third one. The first dynamical landscape, in light red, corresponds to the Hot-Warm House time interval. The second one, in light blue, represents the Cold-Ice house time interval. The third one, in light green, highlights the potential new dynamical landscape represented by the Anthropocene time interval. The different abrupt transitions identified in the present study are reported as TPos or RTPos to distinguish the major tipping points up to the early Quaternary period from the more recent TPs characterizing less drastic climate changes during the Quaternary. Various plate tectonic and ice sheet events are indicated and supported by maps of plate movements and North and South Hemisphere ice sheets. The Antarctica maps are from Pollard & DeConto[79], Northern Hemisphere ice sheet maps are from Batchelor et al.[80]. The paleogeographic maps have been generated using the Ocean Drilling Stratigraphic Network (ODSN) plate tectonic reconstruction service: <https://www.odsn.de/odsn/services/paleomap/paleomap.html>. The red arrows on the tectonic maps indicate the key events that correspond to the identified abrupt transition.

More recently, variations in the extent and volume of the NHIS have contributed to the occurrence of the millennial variability marked by the Bond cycles, which have been best described during the last climate cycle. However, the onset of these cycles has been proposed to date back to 0.9 Ma[72]. Human activity is now rapidly pushing the Earth system towards the limits of its safe operating space associated with the occurrence of TPs; see also Rothman[9] for a geochemical perspective focusing on the ongoing perturbation imposed on the marine carbon cycle.. This concern is supported by actual observations impacting numerous tipping elements (see Lenton[14]) through very drastic tipping cascades[16]. This paves the way for a possible upcoming major transition, which might lead us to climate conditions that are fundamentally different from what has been observed in the recent or more distant past, and, at the very least, could bring us into a climate state with a much reduced or absent NHIS[38]. This potential major transition, leading to *de facto* irreversible changes for the climate and the biosphere, could become a boundary between the Cenozoic icehouse and a new, warmer and radically different climate state compared to Pleistocene conditions.

**Methods**

**Recurrence Plots and Kolmogov-Smirnov Test**

The Kolmogorov-Smirnov test is a robust method for accurately detecting discontinuities in a particular time series and is therefore a very precise way for determining the timing of abrupt transitions. Our method, described by Bagniewski et al.[26] is modified from the two-sample KS test and has been successfully applied to various geological time series[24-25,69]. The method uses the KS statistic, $D_{KS}$, to compare two sample distributions taken before and after a potential transition point within a sliding window. If the samples do not belong to the same continuous distribution, i.e., $D_{KS}$ is greater than a predefined threshold, a transition point is identified.

This classic two-sample KS test is augmented by additional criteria to refine the results and pinpoint significant abrupt transitions indicative of a true climatic shift. This involves discarding transitions below a rate-of-change threshold since smaller changes in the time series might be due to data errors or short-term variability within intervals shorter than the proxy record's sampling resolution. Furthermore, as the frequency with which the KS test detects transitions largely depends on the chosen window length, $D_{KS}$ is calculated for different window length, within a range that corresponds to the desired time scale at which a given paleorecord is to be investigated. The analysis starts by identifying transitions using the longest window, which has the largest sample size and thus carries greater statistical significance. Subsequently, the method incorporates transitions detected using shorter windows to capture transitions occurring on shorter time scales. For more details see Bagniewski et al.[26], Rousseau et al.[72].

To gain further insight into the climate story revealed by proxy records, we performed an analysis based on recurrence plots (RPs), first introduced by Eckmann[81] in the study of dynamical systems and later popularized in the climate sciences by Marwan et al.[53,82]. The RP for a time series $\{x_i : i = 1, ..., N\}$ is constructed as a square matrix in a Cartesian plane with one copy $\{x_i\}$ of the series on the abscissa and another copy $\{x_j\}$ on the ordinate, with both axes representing time. A dot is entered into a position (i, j) of the matrix when $x_j$ is sufficiently close to $x_i$. For the details — such as how "sufficiently close" is determined — we refer to Marwan et al.[50].

The square matrix of dots that is the visual result of RP exhibits a characteristic pattern of vertical and horizontal lines, indicating recurrences. These lines sometimes form clusters that represent specific periodic patterns. Eckmann[81] distinguished between large-scale *typology* and small-scale *texture* in the interpretation of RPs. The most interesting typologies in RP applications are associated with recurrent patterns that are not strictly periodic and are thus challenging to detect by purely spectral approaches to time series analysis. RPs offer an important advantage by enabling a visual identification of nonlinear relationships and dynamic patterns that characterize high-dimensional systems subject to time-dependent forcing, such as the climate system. However, in periodically forced systems, the recurrence structure is dominated by repetitive, periodic patterns rather than the underlying dynamics. This may pose challenges when attempting to identify meaningful transitions or regime shifts in climate at time scales strongly affected by orbital forcing.

Marwan et al.[82] extensively discussed the various measures used to objectively quantify the typologies observed in RPs. Collectively known as recurrence quantification analysis (RQA), these measures include the Recurrence Rate (RR), which represents the probability of a specific state recurring within a given time interval. The RR is calculated by determining the density of recurrence points along the diagonal of the RP within a sliding window. Since low RR values correspond to an unstable behavior of the system, the minima of the RR may be used to identify abrupt transitions. Here we follow the approach by Bagniewski et al. [26] and select local minima based on their prominence.

**Quasi-potential, Melancholia States, and Critical Transitions**

Traditionally, tipping points are schematically represented as being associated with the bifurcation occurring for a system described by a one-dimensional effective potential when a change in the value of a certain parameter leads to a change in the number of stable equilibriums. Hence, conditions describing the nearing of a tipping point can be related to the presence of slower decay of correlations (critical slowing down). This viewpoint, while attractive, suffers from many mathematical issues due to the fact that the true dynamics of the system occurs in a possibly very high dimensional space. Tantet et al. [83,84] have introduced a mathematically rigorous framework for the occurrence of tipping points that clarifies the link between rate of decay of correlations, sensitivity of the system to perturbations, and robustness of the unperturbed dynamics.

Here, we wish to take a different angle on the problem. Instead of focusing on the individual tipping points, we attempt to capture the global stability properties of the system. We take inspiration from the application of the Waddington epigenetic landscape to describe morphological evolution[42–46] and from the theory of punctuated equilibrium[49,50], which associates periods of stasis (characterized by relatively stable morphology) with the emergence of new species through abrupt changes (named cladogenesis) from a previous species. Lucarini & Bodai[39,40] proposed describing the global properties of the climate system using the formalism of quasi-potential[41]. Roughly speaking, assuming that the system lives in $\mathbb{R}^N$, and its dynamics is described by a stochastic differential equation the probability that its state is within the volume $d\vec{x}$ around the point $\vec{x} \in \mathbb{R}^N$ is given by $P(\vec{x}, d\vec{x}) = \rho(\vec{x})d\vec{x}$ where $\rho(\vec{x}) \approx e^{-\frac{2\Phi(\vec{x})}{\varepsilon^2}}$ is the probability distribution function (pdf) and $\Phi(\vec{x})$ is the quasipotential.

The function $\Phi(\vec{x})$ depends in a nontrivial way on the drift term and noise law defining the stochastic differential equation. This setting generalizes the classical energy landscape and applies to a fairly large class of stochastic dynamical systems. One can see the dynamics of the system as being driven towards lower values $\Phi(\vec{x})$ (plus an extra rotational effect that is typical of non-equilibrium systems), while the stochastic forcing noise makes sure that the system is erratically pushed around. Hence, the minima of $\Phi(\vec{x})$ correspond to local maxima of the pdf, and the saddles (which coincide for both $\Phi(\vec{x})$ and $\rho(\vec{x})$) coincide with the Melancholia (M) states[34]. Such M states are unstable states of the system that live at the boundary between basins of attraction and are the gateways for the noise-induced transitions between competing stable states. Margazoglou et al. [37] applied this method to the investigation of the metastability properties of an intermediate complexity climate model and suggested that the presence of decorations of the quasi-potential at different scales could be interpreted as being associated with a hierarchy of tipping points. Indeed, passing near M states is intimately associated with the occurrence of critical transitions. Hence, the construction of the quasi-potential $\Phi(\vec{x})$ can be seen as the structural counterpart of the investigation of the time-evolution of the system and of its critical transitions. Zhou et al. [85] give a complete overview of different methods applied to perform such an analysis.

**Acknowledgments**

We would like to thank our colleagues from Horizon 2020 TiPES project (grant No. 820970) and especially Michael Ghil for useful discussions about this study. This is LDEO contribution and TiPES contribution XXX. VL acknowledges useful exchanges with T. Bodai, R. Boerner, R. Deeley, G. Margazoglou, C. Nesbitt, and L. Serdukova. We would like to thank the editor and reviewers' feedback, which has helped us to improve the clarity and quality of our work.

**Author contributions:** Conceptualization: DDR; Methodology, investigation and Visualization: DDR, VL, WB; Writing—original draft: DDR; Writing—review & editing: DDR, VL, WB.

**Data and materials availability:** All data generated by the present study from the main text or the supplementary materials will be submitted to PANGAEA data repository. U1308 marine data are available at https://doi.org/10.1594/PANGAEA.871937 (Hodell and Channell, 2016b). CENOGRID data are available at https://doi.org/10.1594/PANGAEA.917503 (Westerhold, 2020).

**Additional information**

**Funding:** This research has been supported by the European Commission, Horizon 2020 Framework Programme (TiPES, grant no. 820970), by the Marie Cure ITN Critical Earth (Grant No. 956170) and from the EPSRC Project No. EP/T018178/1
**Competing interests:** All authors declare they have no competing interests.

**Figures legends**

**Figure 1.** KS test and Recurrence Quantification Analysis (RQA) of CENOGRID benthic $\delta^{18}O$. a) Time series in Ma BP with difference of the reconstructed and present Mean Global Temperature in pink). KS test identifying abrupt transitions towards warmer conditions in red and cooler or colder conditions in blue; b) Recurrence plot (RP) with identification of the main two clusters prior and after 34 Ma. The main abrupt transitions identified are highlighted by red circles, and c) Recurrence rate (RR). The pink crosses and vertical green lines indicate the abrupt transitions (TP) detected by the RQA. CENOGRID benthic $\delta^{18}O$ data are from Westerhold et al.[24]

**Figure 2.** Variation through time of three main climate factors and comparison with the identified abrupt transitions (TP) in the CENOGRID benthic $\delta^{18}O$. a) Global Mean Sea Level in meters from Miller et al.[58]. Identification of particular warm and of glaciation events. The Laurentide, GIW-WAIS and Ice free lines are from Miller et al.[58]; b) Carbonate Compensation Depth (CCD) in meters from Pälike et al.[64]. The purple circles identify the TPs on this record; c) Estimate of the $CO_2$ concentration in parts per million volume (ppmv) from Beerling & Royer[65]. The Antarctica glaciation threshold at 750 ppmv and the NH glaciation threshold at 280 ppmv lines respectively are from DeConto et al.[66].

**Figure 3.** Probability density of the climate system in the projected CENOGRID benthic $\delta^{18}O$ and $\delta^{13}C$. space. a) Chronologically ordered $TP_O$s (diamonds) selected according to the KS methodology for $\delta^{18}O$ with time window 1-4 My are shown. The two extra $TP_O$s found via RP are indicated with a *. b) Chronologically ordered fast $TP_O$s ($FTP_O$s, diamonds) selected according to the KS methodology for $\delta^{18}O$ with time window 0.25-1 Myr are shown. c) Same as a), but for the $\delta^{13}C$ record. The approximate timing of the TPs is indicated (rounded to .01 My). The 5 Ky-long portions of trajectories before and after each TP are also plotted.

**Figure 4.** RQA of U1308 benthic $\delta^{18}O$. a) Time series in ka BP; b) RP; and c) RR. Pink crosses and green lines as in Fig. 1. TPo9-10 and RTPo1-4 abrupt transitions are identified from the RR. U1308 benthic $\delta^{18}O$ data are from Hodell & Channell[25]

**Figure 5.** Evolution of the Earth Climate history among 2 different tipping dynamical landscapes and proposal for a potential third one. The first dynamical landscape, in light red, corresponds to the Hot-Warm House time interval. The second one, in light blue, represents the Cold-Ice house time interval. The third one, in light green, highlights the potential new dynamical landscape represented by the Anthropocene time interval. The different abrupt transitions identified in the present study are reported as TPos or RTPos to differentiate the major tipping points from the critical transitions characterizing transitions of lighter significance in the climate history. Various plate tectonic and ice sheet events are indicated and supported by maps of plate movements and North and South Hemisphere ice sheets. The Antarctica maps are from Pollard & DeConto[79], Northern Hemisphere ice sheet maps are from Batchelor et al.[80].

The paleogeographic maps have been generated using the Ocean Drilling Stratigraphic Network (ODSN) plate tectonic reconstruction service: <https://www.odsn.de/odsn/services/paleomap/paleomap.html>. The red arrows on the tectonic maps indicate the key events that correspond to abrupt transition.

# A Punctuated Equilibrium Analysis of the Climate Evolution of Cenozoic exhibits a Hierarchy of Abrupt Transitions

Denis-Didier Rousseau, Witold Bagniewski, Valerio Lucarini

# Appendices

**A Historical Account of the Critical Transitions**

In addition to the Chicxulub meteor impact, which injected a considerable amount of $CO_2$ into the atmosphere[1,2], Deccan traps were already spreading at the beginning of the Cenozoic, contributing to the release of massive amount of $CO_2$[3]. $CO_2$ concentrations continued to rise until reaching about 500 ppmv, coinciding with a very active period of the North Atlantic Igneous Province around 58 Ma – 56 Ma ($TP_O2$ and $TP_O3$), which was associated with the opening of the North Atlantic Ocean[4]. During this time, the Northern Hemisphere plates were connected and did not experience the present Arctic conditions, allowing faunal and vegetal dispersion. In contrast, other plates were undergoing reorganization, such as India moving northeastward towards the Asian continent. Equatorial Pacific carbonate compensation depth (CCD) reached its minimum value a bit later at about 54 Ma when $CO_2$ concentrations were at a maximum for the whole Cenozoic, above 1,100 ppmv[5]. The decrease in GMSL and the CCD at $TP_O4$ (about 40 Ma) has been interpreted as the start of the icing of Antarctica through dating mountain glacier deposits using K-Ar dating of lava flows[6], as well as other glacial evidence, i.e. from the Gamburtsev subglacial mountains[7]. During this period, the Northern Hemisphere plates remained connected. By $TP_O4$ India was approaching the Asian plate while Northern Hemisphere plates remained connected, allowing continental migrations of mammals at high latitudes[8,9]. Around 34 Ma, the EOT ($TP_O6$) is associated with the opening (in several steps) of the Drake and Tasmanian passages[10], which led to a drastic change in the global ocean circulation. This resulted in a reduced strength of deep water formation in the Southern Hemisphere, and a decline in deep-sea temperatures, accompanied by a significant drop in relative sea level and a decrease in CCD (see Fig. 2). Based on paleoaltimetry estimates using oxygen isotopes, Rowley & Currie[11] indicate that the Tibetan Plateau had an elevation of about 4000m, favoring therefore the physical weathering of rocks, resulting in the consumption of $CO_2$, and the enhanced burial of carbon through high sedimentation rates in the nearby seas. This may have contributed to the major threshold in the variation of the $CO_2$ concentration[12], which also drops very strongly[5] (Fig. 2).

Recent analysis of the evolution of sea surface temperatures in the North Atlantic Ocean during the Eocene-Oligocene Transition (EOT) suggests that the cooling during this period was primarily triggered by a decrease in CO2 concentration, with paleogeographic changes playing a secondary role[13]. Comparisons between marine and terrestrial data and models reveal a more abrupt decrease in pCO2 in marine proxies, as evidenced by the decrease in the CCD, contrasting with a more gradual decrease in terrestrial indicators. The gradual decrease observed in terrestrial proxies remains a topic of debate. Additionally, these studies suggest that paleogeographic changes played a significant role in initiating the Atlantic Meridional Overturning Circulation (AMOC), which became an important ocean circulation pattern in the growth of polar ice sheets (Hutchinson et al.[14], and references herein).

As mentioned, The EOT transition is the major boundary between two different climate landscapes dominated by intensive plate tectonics and strong volcanism for the older one, and by major ice sheet coverage in both Hemispheres, with still very active plate tectonics (closure of seaways, orogenies), for the younger one[9]. Following the EOT, the climate experienced the build-up of the East Antarctic ice sheet (Oi-1 glaciation), which can be regarded as the onset of the cold world in which we are presently living. During this time, India has almost ended its transfer to the Asian plate. Between $TP_O$ 6 and $TP_O$ 7, i.e., 34 Ma and 14 Ma respectively, the East Antarctic ice sheet underwent waxing and waning with several major glaciations occurring before the 17 Ma to 14.5 Ma interval, characterized by rising sea levels, a severe shoaling episode of CCD, and higher $CO_2$ concentrations (Fig. 2). Such high $CO_2$ concentrations may have been fueled by the Columbia River major volcanism, which ended by $TP_O$ 7 at about 14 Ma[3]. Plate tectonics still played a very active role, with the closure of both the Indonesian gateway and the Tethyan seaway, and contributing to the start of the development of the Mediterranean [75]. Eurasia is now separated from North America and Greenland, India colliding with the Asian continent, and the Andes are uplifting, thus modifying the geometry of the marine basins and the global oceanic circulation. West Antarctica is beginning to form while the East Antarctic ice sheet continues to strengthen and expand. $TP_O$8 at about 9 Ma sees a strong lowering of the GMSL and of the $CO_2$ concentrations (Fig. 2). $TP_O$9 is associated with a final major tectonic event corresponding to the closure of the Panama Isthmus, which, as a result of shutting down the exchange of water between the two oceans, led to re-routing large-scale flows in both the Pacific and the Atlantic Ocean, and configured a oceanic circulation that is very similar to present day[15]. This is associated with a significant decrease in the global sea level, a deepening of the CCD, and lower $CO_2$ concentrations (Fig. 2). These conditions will shape the Earth climate history during the Quaternary, leading to the build-up of the Northern Hemisphere glaciers and ice sheets. It is worth noting that all the major transitions that have been identified during the interval between 66 Ma and 2.9 Ma - 2.5 Ma are associated with the build-up, waxing, and waning of the Northern and Southern Hemisphere ice sheets.

# A Punctuated Equilibrium Analysis of the Climate Evolution of Cenozoic exhibits a Hierarchy of Abrupt Transitions

Denis-Didier Rousseau, Witold Bagniewski, Valerio Lucarini

# Supplementary Information

**Comparison between GMSL, CCD and $CO_2$ concentration from CENOGRID (past 66 Myr)**

We provide here a more detailed discussion of $TP_O$ 1-9 in relation to reconstructed global mean sea level (GMSL), Pacific carbonate compensation depth (CCD), and $CO_2$ concentration. The interval between 66 Ma and 63 Ma ($TP_o1$) shows a relative stable GMSL above +60m. It is followed by a lowering by about 70m over several steps between $TP_o1$ at 63 Ma and $TP_o3$ at 56 Ma, punctuated by an abrupt increase of 52m at $TP_o2$ around 58 Ma, and another increase of 28m corresponding to the short but intense Paleocene-Eocene Thermal Maximum (PETM – Fig. 2a) warming. Subsequently, GMSL raises by about +40m between $TP_o2$ at 58 Ma and 55 Ma. The interval between 55 Ma and 48 Ma, the Early Eocene climatic optimum (EEOC – Fig. 2A), is characterized by a relatively high GMSL of about 66 m above the present mean sea level, which is associated with the occurrence of hyperthermal conditions at 58 Ma, 57 Ma, and 53 Ma. Between 52 Ma and TP4 at about 47 Ma, CCD reaches the shallowest depth of the record, about 3000m [1], associated with the highest $CO2$ concentrations estimated above 1100 ppmv by Beerling and Royer [2] – Fig. 2B,C). A strong decrease in GMSL of about 30m occurs between 48 Ma and 46 Ma while it remains relatively stable at about +42m between 46 Ma and TP5, at 40 Ma. It is however punctuated by two short events, first a lowering of about 25m between 42 Ma and 40.5 Ma, and next by a strong increase of about 40m between 40.5 Ma and 40 Ma ($TP_o5$) corresponding to the Middle Eocene Climatic optimum (MECO – Fig. 3A). A two-step lowering of about 35m of the GMSL occurs between 40 Ma ($TP_o5$) and 36 Ma, followed by a gradual increase of about 25m between 36 Ma and 34.5 Ma just before the Eocene-Oligocene Transition (EOT – Fig. 3A). Between 46 Ma and 34 Ma ($TP_o6$), the CCD is strongly oscillating with numerous deepening and shoaling events of 500 to 1000 m in magnitude and shoaling ones, corresponding to carbonate accumulation episodes. These oscillations in the CCD occurred during an interval indicating still high $CO2$ concentrations, roughly above the 750 ppmv considered as the Antarctic glaciation threshold [3], and marked by a relative minimum at about $TP_o5$ at 40 Ma (Fig. 3B,C). This first interval of the GMSL, which ends with a strong deepening of about 1000m, is associated with a strong decrease in $CO_2$ concentration and an almost completely ice free Earth, with no major ice sheet in either the southern or northern hemisphere. This is deduced primarily from the high GMSL, mostly remaining above 12m (Fig. 2A), which would correspond to the mutual contribution of the Greenland and the West Antarctic ice sheets. Additionally, high $CO_2$ concentrations estimates are in agreement with a CCD generally lower than 4000 m (Fig 2B,C).

The second main interval shows a completely different scenario with the GMSL varying between +30m and -80m without considering the late Quaternary interval, and much lower CCD and CO2 concentration (Fig. 2B,C). First the strong decrease in the GMSL at EOT reaches negative values of about -25m at about 33.5Ma. It is interpreted as evidence of the first continental scale Antarctic ice sheet (first Oligocene isotope maximum - Oi1 – Fig. 2A) [4]. After a return to values similar to present mean sea level, about +2m, between 32 Ma and 30 Ma, a new decrease of about 25m in GMSL occurs between 29.5 Ma and 27 Ma corresponding to another continental scale Antarctic ice sheet extent labeled Oi2 (Fig. 2A). After a two-step increase in GMSL of about 40m between 27 Ma and 24 Ma, a new sharp decrease of about 40m is noticed at about 23 Ma. It corresponds to the Middle Oligocene Maximum (Mi1 – Fig. 2A), another Antarctic ice sheet wide expansion [4,5]. From $TP_o6$, at 34 Ma, and 23 Ma, the $CO_2$ concentration decreases associated with a deepening trend in the CCD down to about -4600m. From 23 Ma until about 19 Ma, GMSL shows oscillations but with lower values than present day at about -20m, whereas from 19 Ma until 17 Ma, GMSL increases by about 50m to indicate high values around +30 m above present day value (Fig. 2A). The CCD indicates about 600 m shoaling which lasted around 2.5 Myr linked to high estimates of $CO_2$ concentration from paleosols and stomata [2] (Fig. 2B,C). This strong increase corresponds to the Miocene Climatic optimum (MCO – Fig. 2A) between 17 Ma and $TP_o7$ at 13.9 Ma, which is the last interval during which GMSL reaches such high values higher than +20m above the present day ones (Fig. 2). The interval between $TP_o7$ at 13.9 Ma and 13 Ma corresponds to the Middle Miocene transition during which GMSL once more decreases significantly by about 35m. Such lowering is associated with the growth of the East Antarctic ice sheet to near its present state, remaining a perennial ice body that is thereafter impacting the Earth climate [6,7]. Although GMSL remains relatively stable between 13 Ma and 12 Ma, another strong decrease, again of about 30m, occurs between $TP_o8$ at about 9 Ma and 8.5 Ma, associated with the strongest deepening recorded by the CCD, around 4800m. GMSL increases again by about 20m until 7.5 Ma to remain relatively stable until 5.5 Ma when GMSL increases by about 20m between 5.5 Ma and 3.5 Ma, corresponding to the Pliocene Climatic Optimum (PCO – Fig. 2A). Between 3.5 Ma until $TP_o9$ at about 2.7 Ma GMSL shows a sharp decreasing trend of about 35m (Mi2a, 3, 4 – Fig. 2A). This is associated with the development of the large northern ice sheets between 2.9 Ma ($TP_o9a$) and 2.5 Ma ($TP_o9b$ – Fig. 2), especially the Laurentide ice sheet corresponding to about 50m decrease of GMSL with regards to the present day value (Fig. 2A). From $TP_o8$ at about 9 onward, CCD shows significant fluctuations, although it remains at around 4500m with two strong deepening events at about $TP_o9a$ 2.9 Ma and $TP_o9b$ (Fig. 2) at 2.5 Ma.

In another global sea level reconstruction[8], several major thresholds were identified agreeing with the Miller et al [10] reconstruction and our present analysis. This reconstruction estimates the EOT global sea level drop at 34 Ma ($TP_o6$) as about 30m, while previous reconstructions by Houben et al. [9], and by Miller et al. [10] found a decrease of 70m-80m. Miller et al [10] also estimate this drop in global sea level being associated with a 2.5°C cooling interpreted previously as above the onset of the Antarctic ice sheet glaciation. Rohling et al. [8] also identify 14 Ma ($TP_o7$) threshold as the end of the last intermittently ice free period in the Earth history of the last 40 Ma with only the southern hemisphere ice sheets impacting the Earth climate. Miller et al. [10] indicates a 35m lowering at that particular transition. Indeed Rohling et al. [8] indicate a slight negative sea level shift at around 10 Ma ($TP_o8$), of about 10m, which they interpret as the onset of partial or ephemeral northern Hemisphere ice bodies with two other sea level thresholds at about 3 Ma (close to $TP_o9a$) and 2.75 Ma ($TP_o9b$), also observed in Miller et al. [10]. These two key dates correspond to respectively to the first major iceberg calving in the Nordic Seas [11] and from the Laurentide ice sheet [12] although this interpretation was rejected by Naafs et al. [13] who instead attributed the IRD signature to Greenland and Fennoscandian glaciers. Hodell and Channell [14] identified the first occurrence of iceberg calving in North Atlantic at about 2.75 Ma from the analysis of the $\delta^{18}O$ of benthic bulk carbonate in core U1308 (Supp. Fig. 3), a date identified as an abrupt transition in the RR analysis of both $\delta^{18}O$ of benthic foram and bulk carbonate [15]. All along the past 66 Ma, GMSL has been varying between average values of +38. m ± 15 m above the present day value during the hot world interval prior to $TP_o6$ at 34 Ma, and of -3.5m ± 13 m from 34 Ma until the present day during the cold world interval (Tab. 2).

**The past 3.3 Myr.**
We provide here a more detailed discussion of the past 3.3 Myr. The first date, $TP_o9b$ detected by RQA, is interpreted as corresponding to the earliest occurrence of IRD in the North Atlantic. This occurrence characterizes the presence of Northern Hemisphere coastal glaciers large enough to calve icebergs in the ocean, and the melting of these icebergs is likely to have impacted the oceanic circulation. Naafs et al. [13], however, reported the occurrence of weak IRD events in the late Pliocene that they attributed mainly to Greenland and Fennoscandian glaciers. Nevertheless, such interpretation points to nevertheless smaller ice sheets over these regions than during the later Quaternary, when North American ice sheets were considerably larger. The interval $TP_o9a$, at 2.8 Ma, to $RTP_o2$, at 1.2 Ma, shows glacial–interglacial sea level variations of about 25–50 m below the present day. The CO2 concentrations varied between 270 ppmv and 280 ppmv during interglacials and between 210 ppmv and 240 ppmv during glacials, with a decreasing trend of about 23 ppmv over this 1.4-Myr–long interval [16].

The second date, $RTP_o1$, at 1.55 Ma, corresponds to an increased amplitude in ice volume variations between glacial minima and interglacial optima. This second step shows the permanent occurrence of ice-rafted events during glacial intervals in the record (Suppl. Fig. 3), therefore indicating an amplified relationship of climate variations with Northern Hemisphere ice sheets. The increase in IRD variability and magnitude since $RTP_o1$, however, shows that distinct, faster processes have to be considered than those due to slow changes in Earth's orbital parameters; see again Fig. 4.

The third date, $RTP_o2$, at 1.25 Ma, close to the MIS22–24 $\delta^{18}O$ optima, shows increased continental ice volume in the Northern Hemisphere [17], but also more stability in the East Antarctic ice sheet in the Southern Hemisphere [18]. In parallel, evidence of a major glacial pulse recorded in Italy's Po Plain, as well as in $^{10}$Be-dated boulders in Switzerland, is interpreted as marking the onset of the first major glaciation in the Alps [19,20].

After RTP$_o$2, at 1.25 Ma, the sea level decreased to about 70–120 m below the present day, while the CO2 concentrations varied between 250 ppmv and 320 ppmv during interglacials and between 170 ppmv and 210 ppmv during glacials [21]. Similar variations were determined by Seki et al. [22], although pCO2 changes that occurred before the time reached by ice core records are associated with high uncertainties in both dating and values. The sawtooth pattern of the interglacial– glacial cycles [23] becomes noticeable at 0.9 Ma. At about the same time, the synthetic Greenland $\delta^{18}$O reconstruction indicates the occurrence of millennial variability expressed by DO-like events [24].

Finally, RTP$_o$3, at 0.65 Ma, marks the end of the transition from the Lower and Mid-Pleistocene interval — characterized by 41-Kyr–dominated cycles and smaller 23-Kyr ones — to the Upper Pleistocene, with its 100-Kyr–dominated cycles; see Fig. 4. The sawtooth pattern of the interglacial–glacial cycles is well established during this final interval, in contradistinction with the previous, more smoothly shaped pattern that appears to follow the obliquity variations. The global ice volume is maximal, exceeding the values observed earlier in the record, especially due to the larger contribution of the Northern American ice sheets. The latter now have a bigger impact on Northern Hemisphere climate than the Eurasian ice sheets [17]. The IRD event intensity and the frequency of occurrence increase [25] as well (Suppl. Fig. 3), leading to the major iceberg discharges into the North Atlantic named Heinrich events (HEs); see [26–29]. The interval of 1 Ma – about 0.4 Ma (RTP$_o$4) is also the interval during which Northern Hemisphere ice sheets reached their southernmost extent [17]. Applying Mg/Ca transfer functions, Elderfield et al. [30] have estimated that the past 0.4 Myr water temperatures have been the highest the past 1.2 Myr, supporting the local temperature variations deduced from the Antarctic ice cores [31]

Figure S1: KS test and Recurrence Quantification Analysis (RQA) of CENOGRID benthic δ13C. A) KS test identifying abrupt transitions towards warmer conditions in red and cooler or colder conditions in blue; B) Recurrence plot (RP), and C) Recurrence rate (RR). The pink crosses and vertical green lines indicate the abrupt transitions (Table) detected by the RQA.

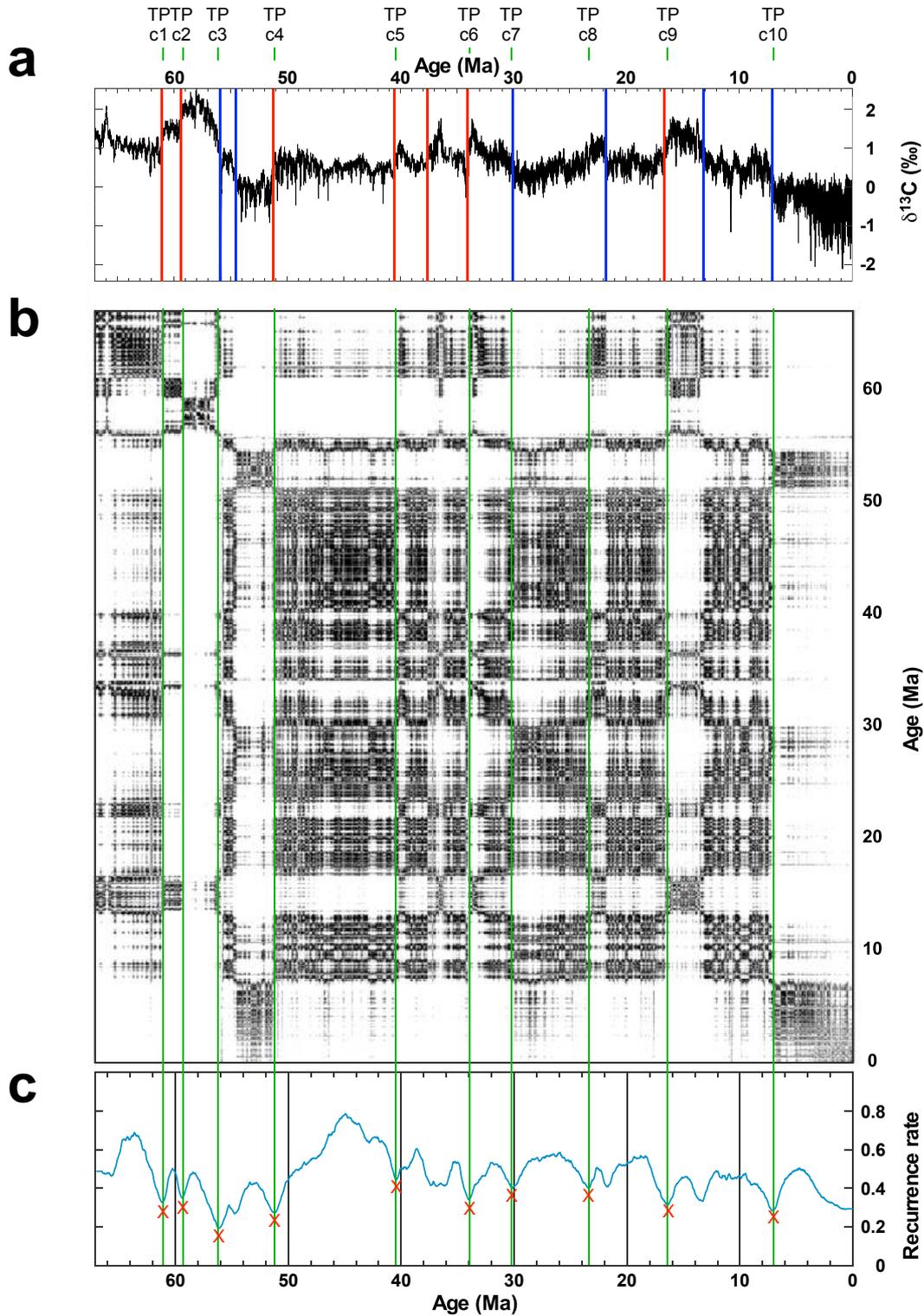

Figure S2: RQA of U1308 benthic $\delta^{13}$C. a) Time series in Ma; b) RP; and c) RR. Crosses as in Fig. S1.

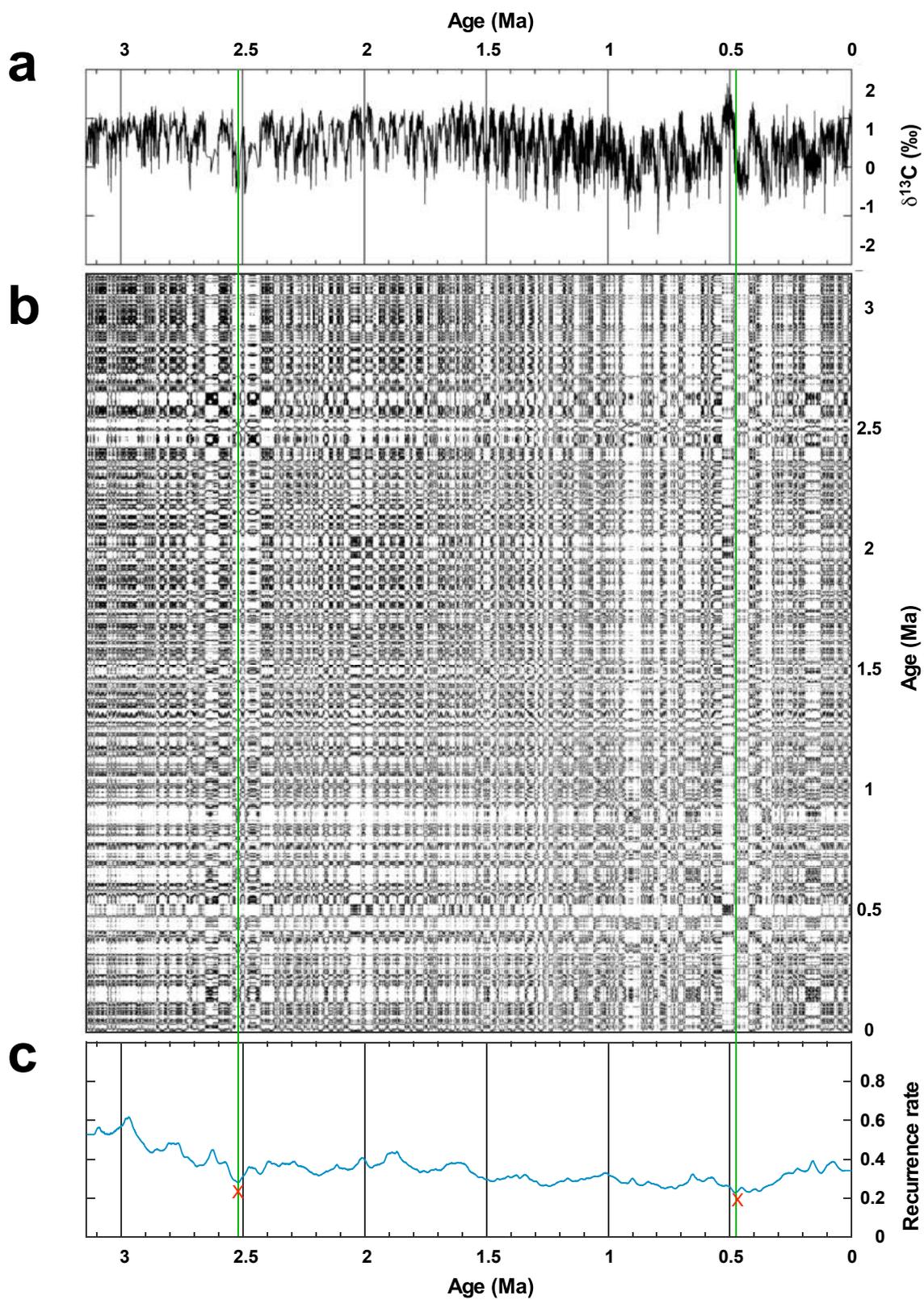

Figure. S3: RQA of U1308 bulk carbonate $\delta^{18}O$. A) Time series in Ma of benthic $\delta^{18}O$ in blue and $\delta^{18}O$ bulk carbonate in green; b) RP; and c) RR). The pink crosses and vertical lines indicate the abrupt transitions (Table) detected by the RQA.

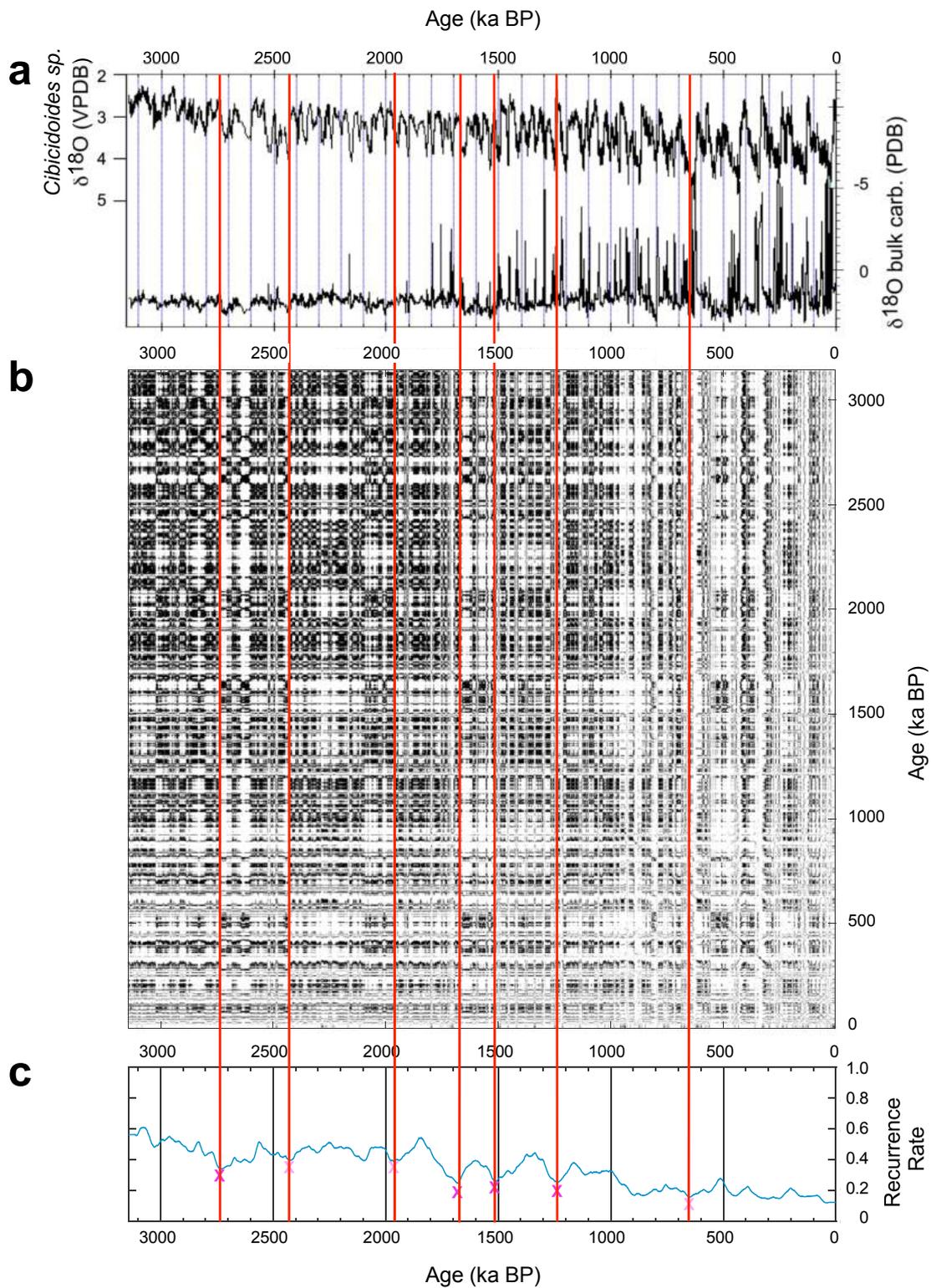

Table S1: Recurrence Quantification Analyses (RQA) of CENOGRID benthic $\delta^{18}O$ and $\delta^{13}C$, and U1308 benthic $\delta^{18}O$, $\delta^{13}C$, and bulk carbonate $\delta^{18}O$. For each, listed are the dates of the abrupt transitions, the corresponding RR prominence values used to select the abrupt transitions, and the significance of those transitions. X denotes prominence values above the RR standard deviation, while $ denotes values below the standard deviation. Ages are in Ma for the CENOGRID and in ka BP for the U1308 records respectively.

| CENOGRID δ18O, window: 1-6 Ma | | | CENOGRID δ13C, window: 1-6 Ma | | | U1308 δ18O, window: 60-250 ka | | | U1308 δ13C, window: 60-250 ka | | | U1308 bulk carbonate δ18O, window: 60-250 ka | | |
|---|---|---|---|---|---|---|---|---|---|---|---|---|---|---|
| Ma BP | RR prominen | Significance | Ma BP | RR prominen | Significance | ka BP | RR prominen | Significance | ka BP | RR prominen | Significance | ka BP | RR prominen | Significance |
| 33,85 | 0,41570578 | X | 56,15 | 0,49582588 | X | 2524 | 0,23796191 | X | 477 | 0,16393608 | X | 2732 | 0,20363551 | X |
| 47 | 0,33114787 | X | 34 | 0,24468463 | X | 1510 | 0,14169216 | X | 2521 | 0,15914759 | X | 1681 | 0,19748231 | X |
| 14,05 | 0,31556165 | X | 7,15 | 0,21940684 | X | 354 | 0,1239728 | X | - | - | - | 1510 | 0,13491356 | X |
| 62,65 | 0,23176521 | X | 61,15 | 0,17631944 | X | 614 | 0,11154923 | X | - | - | - | 1234 | 0,12752732 | X |
| 39,85 | 0,18154617 | X | 16,4 | 0,17583585 | X | 1248 | 0,08692142 | X | - | - | - | 1966 | 0,12498718 | $ |
| 56,05 | 0,17128033 | X | 23,4 | 0,16939465 | X | 2925 | 0,07068202 | $ | - | - | - | 653 | 0,12224033 | $ |
| 58,05 | 0,12919749 | X | 40,45 | 0,16112753 | X | - | - | - | - | - | - | 2421 | 0,11758806 | $ |
| 9,7 | 0,0862254 | $ | 51,2 | 0,15785367 | X | - | - | - | - | - | - | - | - | - |
| - | - | - | 59,4 | 0,13334579 | X | - | - | - | - | - | - | - | - | - |
| - | - | - | 30,2 | 0,13193157 | X | - | - | - | - | - | - | - | - | - |

Table S2: Statistics of the 66-34 Ma and 34 Ma-present intervals, from top to bottom: the Global Mean sea level (GMSL) in meters from Miller et al. [10], the CO2 concentration in ppmv from Beerling and Royer [2], and for the CCD depth in meters from Palike et al. [1].

| Miller et al. (2020) | Age_cal Ma BP | Sea level (m) | Age_cal Ma BP | Sea level (m) |
|---|---|---|---|---|
| Minimum | 0,98 | -33,00 | 33,68 | -1,90 |
| Maximum | 33,66 | 33,20 | 64,82 | 77,30 |
| Points | 1635,00 | 1635,00 | 1558,00 | 1558,00 |
| Mean | 17,32 | -3,49 | 49,25 | 38,47 |
| Median | 17,32 | -3,80 | 49,25 | 36,00 |
| Std Deviation | 9,44 | 12,90 | 9,00 | 14,85 |

| Beerling & Royer (2011) | Age (Ma) | CO2 (ppm) | Age (Ma) | CO2 (ppm) |
|---|---|---|---|---|
| Minimum | 0,00 | 80,00 | 34,00 | 100,00 |
| Maximum | 33,60 | 1232,00 | 65,00 | 1868,00 |
| Points | 289,00 | 289,00 | 81,00 | 77,00 |
| Mean | 14,29 | 329,81 | 49,17 | 626,96 |
| Median | 14,10 | 271,00 | 54,00 | 574,00 |
| Std Deviation | 8,39 | 164,47 | 9,73 | 311,79 |

| Pälike et al. (2012) | Age (Ma) | eq CCD (m) | Age (Ma) | eq CCD (m) |
|---|---|---|---|---|
| Minimum | 0,00 | 4100,00 | 33,75 | 2800,00 |
| Maximum | 33,50 | 4900,00 | 52,25 | 4300,00 |
| Points | 135,00 | 135,00 | 75,00 | 75,00 |
| Mean | 16,75 | 4586,30 | 43,00 | 3518,67 |
| Median | 16,75 | 4600,00 | 43,00 | 3500,00 |
| Std Deviation | 9,78 | 153,01 | 5,45 | 408,92 |

Table S3: Summary of the GMSL, CCD and CO2 concentration trends at the identified abrupt transitions TP1 to TP10.

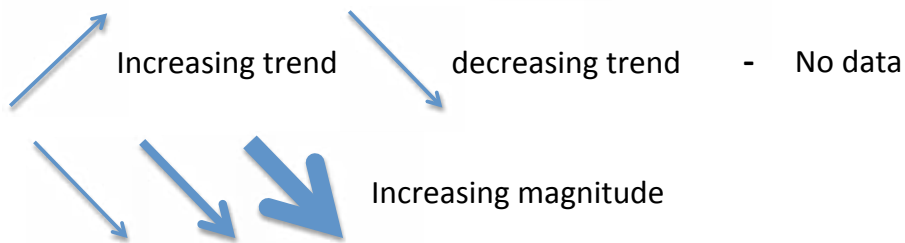